\documentclass[a4paper,aps,pre,twocolumn]{revtex4}
\usepackage{amssymb,amsmath,textcomp,epsfig,txfonts,subfigure}
\usepackage{graphicx}

\begin{document} 
\title{Monitoring spatially heterogeneous dynamics in a drying colloidal thin film}
\author{P. Zakharov$^{(1),*}$ and F. Scheffold$^{(1,2)}$ }
\address{$^{(1)}$Department of Physics, University of Fribourg,
CH-1700 Fribourg, Switzerland; email: Frank.Scheffold@unifr.ch}
\address{$^{(2)}$Fribourg Center for Nanomaterials, University of Fribourg,
CH-1700 Fribourg, Switzerland}
\address{$^{(*)}$present address: Solianis Monitoring AG,
CH-8050 Zurich, Switzerland}
\date{\today}

\begin{abstract}
We report on a new type of experiment that enables us to monitor spatially and temporally heterogeneous dynamic properties in complex fluids. Our approach is based on the analysis of near-field speckles produced by light diffusely reflected from the superficial volume
of a strongly scattering medium. By periodic modulation of an incident speckle beam we obtain pixel-wise ensemble averages of the structure function coefficient, a measure of the dynamic activity. To illustrate the application of our approach we follow the different stages in the drying process of a colloidal thin film. We show that we can access ensemble averaged dynamic properties on length scales as small as ten micrometers over the full field of view. 
\end{abstract}

\maketitle 

\section{Introduction}
Soft materials are frequently studied with traditional scattering techniques using light, neutron or X-ray radiation \cite{Lindner2002}. These techniques provide invaluable information about the structure and dynamics in complex systems spanning timescales from nanoseconds to hours and length scales from nanometer to mm. Scattering methods usually provide information about time or ensemble averaged bulk properties. The method therefore implicitly assumes that the dynamic properties are stationary and fluctuations are Gaussian.  The situation is qualitatively different in many dense soft materials such as gels,
foams, emulsions, pastes, hard-sphere glasses \cite{hohler97:periodic,Lemieux1999,kegel00:direct,weeks00:3d,knaebel:aging,cipelletti:trc,cipelletti05:slow,berthier05:direct,narita04:drying,dawson02:glass,scheffold03:light} that display ultraslow relaxation
processes. A large number of studies have addressed this problem over the last decade. It has been found that slow relaxations are frequently associated with sudden intermittent changes \cite{cipelletti:trc,cipelletti05:slow}. Intermittent events however are not revealed by recording the time
averaged intensity correlation function $g_{2}(\tau)$ . High order correlation functions still require time-averaging but do in fact contain information about non-stationary
dynamics as shown by Lemieux and Durian \cite{Lemieux1999}. The recently introduced time resolved correlation (TRC) approach of Cipelletti and Trappe solves the time averaging problem by using an area detector. The method of TRC determines the correlation coefficient $c(t,\tau)$ by multiplying two far-field speckle images taken at time $t$ and $t+\tau$ without any need for further time averaging \cite{cipelletti:trc}. It is now possible to monitor the \emph{amount} of correlation
$c(t,\tau)$ as a function of correlation time $\tau$, as it is done in a traditional experiment $\langle c(t,\tau) \rangle_t =g_{2}(\tau)-1 $, or as a function of $t$. The advent of this new technique has revealed that intermittency is intimately related to the slow glassy dynamics observed in many dense complex systems \cite{cipelletti:trc,cipelletti05:slow}.

In this article we report on a new echo speckle imaging (ESI) technique that allows to monitor both spatially \emph{and} temporally heterogeneous dynamic properties. This is of particular importance since in many cases the observed intermittency of the relaxation process can be related to a succession of spatially localized events \cite{berthier05:direct}. Experimental evidence for this scenario is however scarce \cite{kegel00:direct, weeks00:3d}. Our new concept is to study properties of the near-field speckles produced by multiply scattered light diffusely reflected from the superficial volume
of a strongly scattering medium. We use a spinning ground-glass disk to scramble the incident beam which allows us to create a large number of statistically independent optical configurations at each point in the image plane. The high resolution both in time and space distinguishes our ESI approach from other recent developments \cite{Duri2009,Erpelding2008}. To illustrate the application of ESI we follow the different stages in the drying process of a colloidal thin film. In the course of this process the suspended particles undergo a continuous transition from liquid to an amorphous solid driven by the evaporation of the solvent and compaction of the solid
component\cite{brown02:consol,scherer:dr,breugem05:dws}.  Due to the evaporation of water the particle dispersion
concentration increases until the colloids are jammed \cite{dawson02:glass} and subsequently the
remaining water evaporates from the porous solid film~\cite{scherer:dr}.

\begin{figure}
  \center
  \begin{minipage}[tc]{0.85\linewidth}
    \center
   \includegraphics[width=\linewidth,trim=10 50 0 0]{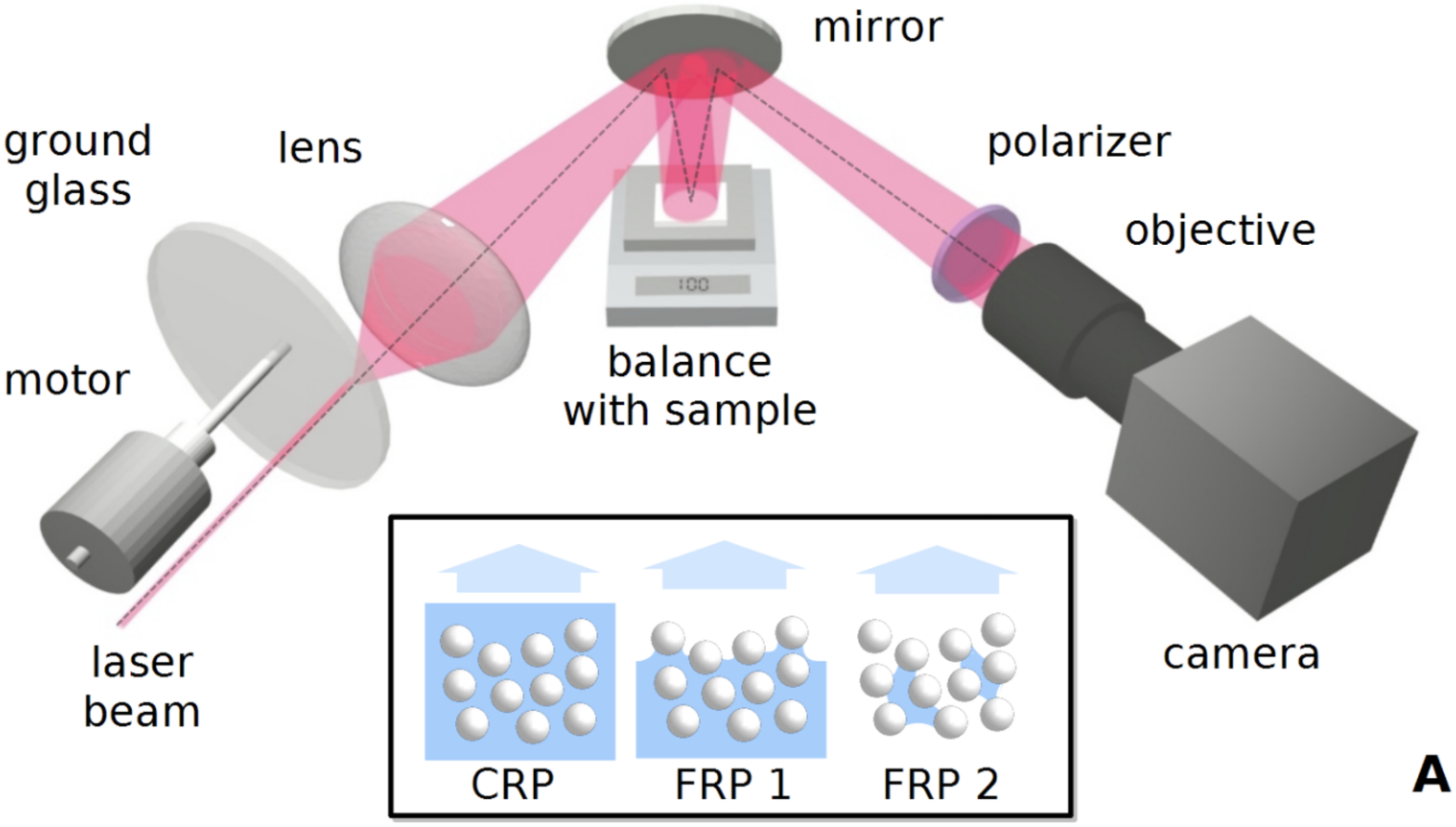}
  \end{minipage}
  \begin{minipage}[tc]{0.85\linewidth}
    \center
     \includegraphics[width=\linewidth,trim=10 50 0 0]{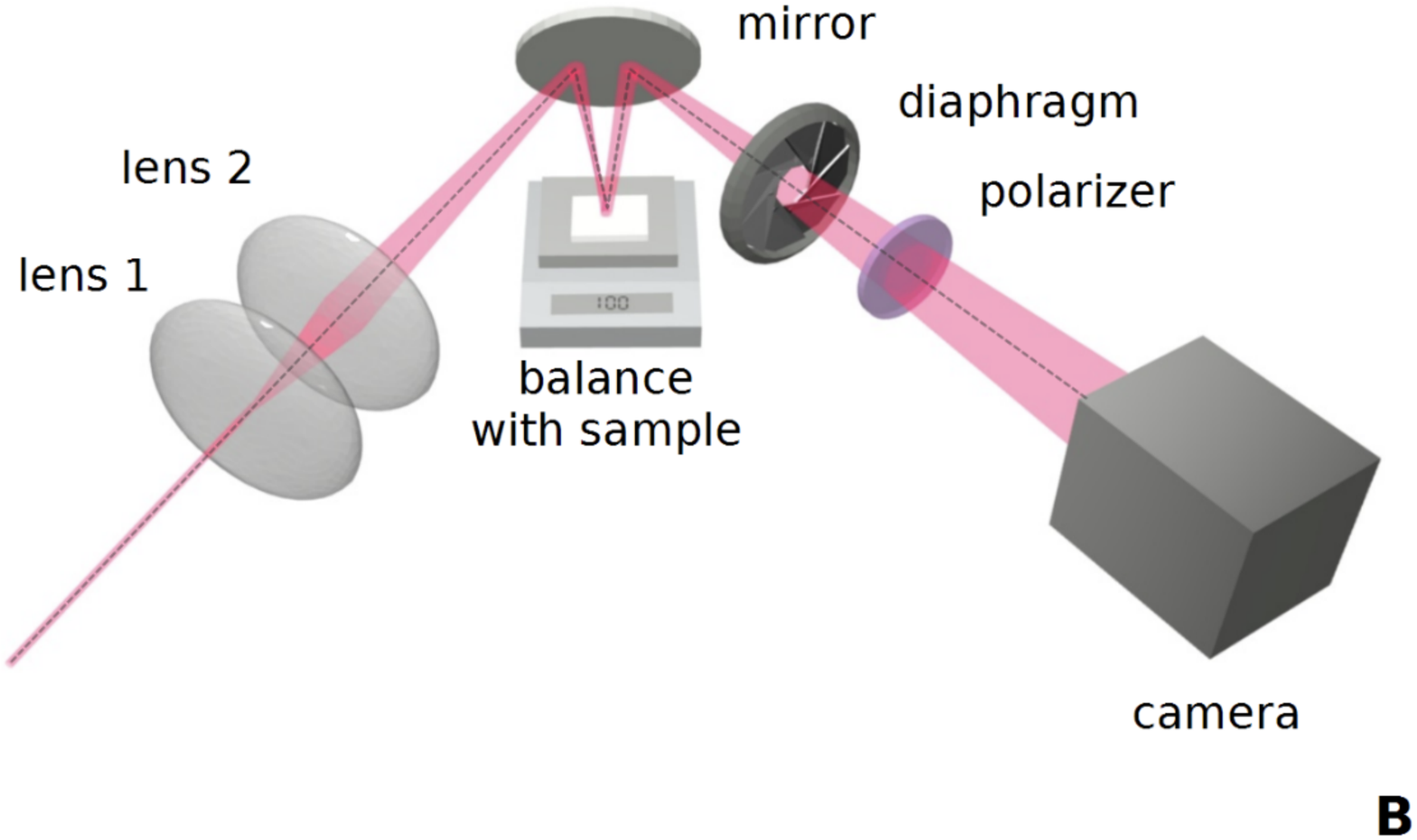}
  \end{minipage}
\vspace{0.1cm} \caption{Experimental set-up: A. Near field laser speckle imaging experiment:  The incident laser
beam is dispersed by a ground glass mounted on a stepper motor. A digital camera records the scattered light (cross polarized channel) in the image plane of the drying paint film. The sample is kept at constant
humidity (65\%) and a balance is used to monitor the weight. The inset shows the main liquid-solid
configurations in the drying process. B. Light scattering experiment in the far-field using a focused incident
beam.} \label{fig:fig1}
\end{figure}

\section{Experiment}
The echo speckle imaging (ESI) approach provides a real-time two-dimensional image of the sample dynamic properties characterized by
the intensity structure function (ISF) using a standard digital camera and imaging optics (Figure \ref{fig:fig1}). Using this method we can access ensemble averaged dynamic properties on length scale as small as ten micrometers over the full field of view 1.6$\times$1.2 mm$^2$. The experiment is set up as follows: light from a HeNe laser (power 15 mW, wavelength $\lambda = 632.8$~nm) is dispersed by a ground glass mounted on a stepper motor. The dispersed light is collimated to create a speckle beam for a homogeneous illumination of the sample surface. The diffuse reflected light is monitored in the image plane with a charge-coupled device (CCD) camera PCO Pixelfly (640$\times$480 pixels of 9.9$\times$9.9~\textmu m$^2$ size, 2$\times$2 pixels binning) at magnification 1.8 positioned at about $50$~cm from the sample. This sets the lateral resolution for a meta pixel (composed of 4 binned physical pixel) to $L_0=11$~micrometers which is larger than the transport mean free path of the diffusely scattered light (in the range 2-5 micrometers) and also large compared to the typical size of the colloidal particles of 0.2-0.5 micrometers. The speckle size is set by the camera diaphragm to about 5-10~$\mu$m. The estimated depth-of-focus of the imaging system is close to 0.1~mm.
The camera is set to acquire 100 images per second with an exposure time of 1~ms.

Sweeping the speckle field over the sample with the stepper motor allows ensemble averaging in a single point on the scattering area. The period is defined by the ground glass rotation frequency of 5~Hz. Correlation echoes after a delay time of $\tau=0.2$~seconds are calculated at each point by comparing 20 sequential images taken at each revolution of the ground glass. Additional time averaging over one second has been performed in order to reduce statistical noise.  We note that this last step could easily be omitted when using a faster digital camera. The sample dynamic properties are analyzed in terms of the normalized ISF:
\begin{equation}
 d_2(t, \tau)= \langle \left [ I \left (t - \tau / 2 \right) - I \left ( t + \tau / 2 \right ) \right ]^2 \rangle /\langle I \rangle^2.
 \end{equation}
Here $\tau$ characterizes the time scale probed and $t$ indicates the time evolution of the experiment \cite{berne76}.
The structure function as a direct measure of the dynamic activity has several advantages over the commonly used intensity correlation function (ICF) $g_2(t,\tau)= \langle \left [I(t - \tau/2)
I(t+\tau/2) \right ] \rangle / \langle I \rangle^2$. While both quantities are directly related in the limit of
perfect measurement statistics $d_2(t, \tau) = 2 \left [g_2(t, 0)- g_2(t,\tau) \right ]$, the ISF is known to outperform the ICF in accuracy when the collection time is limited and further the ISF is less sensitive to low frequency noise or drifts \cite{schatzel:sf}.  Our approach is technically similar to the recently introduced two cell diffusing wave spectroscopy (DWS) echo technique~\cite{zakharov:pre2006,Zakharov2009}. However instead of recording speckle fluctuations in the far-field with a point detector we use a digital camera to monitor fluctuations in the image plane. Due to the fast motion of the ground glass and the finite camera exposure time the experimental values of $d_2(t, \tau)$ are substantially reduced which complicates the normalization of the ISF. Rather than attempting an absolute calibration of $d_2(t, \tau)$ we have chosen to take the initial state of the experiment, $d_2(t=0)$, as a reference. 

\begin{figure}
\includegraphics[width=.75\linewidth]{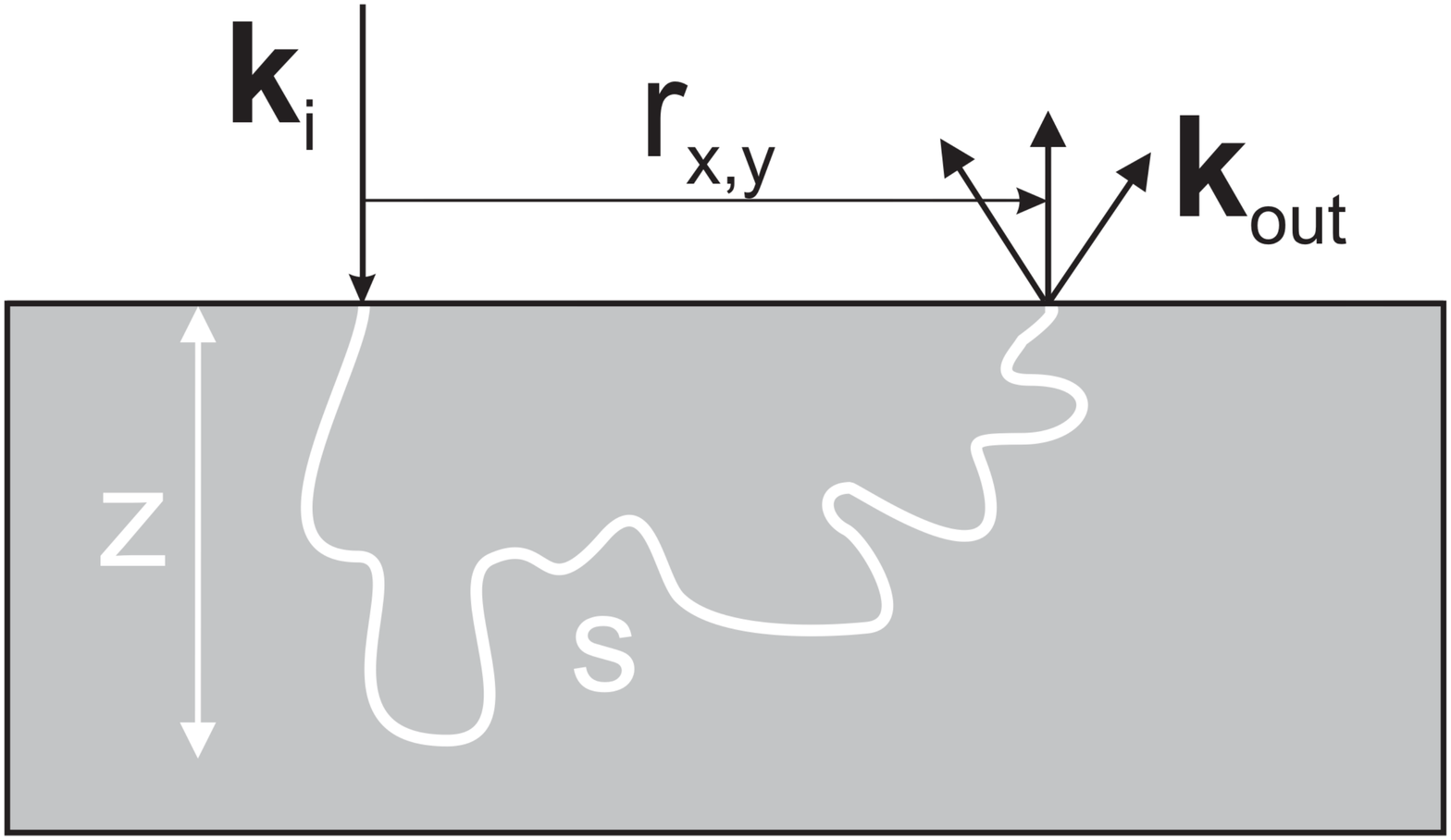}
\vspace{0.1cm} \caption{Multiple scattering of light in backscattering geometry. The incident wave $\bf{k_{in}}$ is dispersed and scattered along a path $s$.  The scattered wave $\bf{k_{out}}$ leaves the sample at a distance $r_{x,y} \le 4l^*$. The propagating waves reach a typical depth $z$ of the same order. Overall the scattered light probes a volume of roughly $3l^* \times 3l^* \times 3l^*$.}
\label{fig0}
\end{figure}

The scattering geometry in our experiment is similar to the one used in traditional backscattering variant of DWS except for the fact that we are studying speckle fluctuations in the image plane \cite{maret1987,pine1988,Goodman2007}. Speckles in the image plane are often referred to as \emph{near field speckles} \cite{Cerbino2009} which must not be confused with the optical near field of evanescent waves probed for example with scanning near field optical microscopy \cite{Betzig1991}. Traditional DWS on the other hand analyzes speckle fluctuations in the far field and therefore does not provide any spatial resolution.  Speckle fluctuations in the near field have been used in biomedical studies of blood flow in the brain cortex or in the eye retina since the early 1980's \cite{briers96:lasca}. This approach is known under the name \emph{laser speckle imaging} (LSI) or \emph{laser speckle contrast analysis} (LASCA) . Without the echo-averaging scheme presented in this article the method is however not sensitive enough in most cases to resolve the subtle differences in the heterogeneous dynamics of soft materials. Only if the speckle dynamics vary rather slowly in space this method can be applied to soft materials. In a recent article Erpelding and coworkers show that the traditional laser speckle imaging approach can be used to characterize the deformation of a soft solid \cite{Erpelding2008}.

Generally speaking the image speckle are formed by multiply scattered light diffusely reflected from the media as illustrated in Figure \ref{fig0}.  A detailed theoretical description of the static scattering process can be found in references \cite{Erpelding2008,Baravian2005,Lenke2000,Haskell1994}. Briefly, the radial distribution of scattered light is approximately given by $I({\bf{r}_{x,y}},s) \simeq \left( {{3 \mathord{\left/ {\vphantom {3 {4\pi sl^* }}} \right. \kern-\nulldelimiterspace} {4\pi sl^* }}} \right)\exp \left[ { - 3r_{x,y}^2 /4sl^* } \right]$, where $l^*$ is a transport mean free path. At the same time the distribution of pathlengths decays as $P(s)\simeq (s/l^*)^{-3/2}$\cite{Lenke2000}. As a result most scattering paths do not extend beyond a few $l^*$.  It can be shown that half of the photons exit the sample at a distance $r_{x,y} \le 2.7 l^*$ \cite{Erpelding2008}. Contributions from single scattering and short paths $s \le l^*$ are suppressed by recording depolarized light only \cite{Rojas2004}. Due to diffuse scattering the length scale probed in $z$ direction is of the same order. Thus at any given point on the sample surface the speckle fluctuations are controlled by the dynamics in a scattering volume of size roughly $\left( {3l^* } \right)^3$. In most cases of practical interest this means that the spatial resolution is set by the transport mean free path of the sample and not by the imaging optics. 

Our sample is a uniformly dispersing white construction paint (Krautol Rollfarbe Super 4062, Germany)
with an initial solid content of 42$\pm$2 vol.~\% in a cell of 200 ~\textmu m thickness and volume 52~\textmu l.
The water based paint contains incompressible TiO$_2$ pigments (typical size $0.2-0.5 \mu$m) as the main light scattering agent and a number of
non-specified, polymer based additives. The upper face of the cell was kept open and the sample surface was
levelled by removing excess paint with a sharp glass plate. A thin polymer ring with inner diameter of 5.5~mm was
immersed in order to prevent the formation of a lateral drying front. The scattering properties of the
paint are found homogeneous over the area of observation.

In order to achieve controlled water evaporation conditions the sample is kept at constant humidity 65$\pm$1~\%
using a home-made computer-controlled optically transparent humidity chamber. Careful precautions were taken in
order to prevent possible air flows from disturbing the measurements which includes porous diffusers for the
humidity controlling fans and an optimized humidity stabilization algorithm. In parallel to the light scattering
measurements the sample weight is monitored by a balance with 1~mg resolution and 1~Hz measurement frequency.
Sub-milligram resolution has been achieved by a polynomial fit to the measured values.

\begin{figure*}
\includegraphics[width=0.85\linewidth]{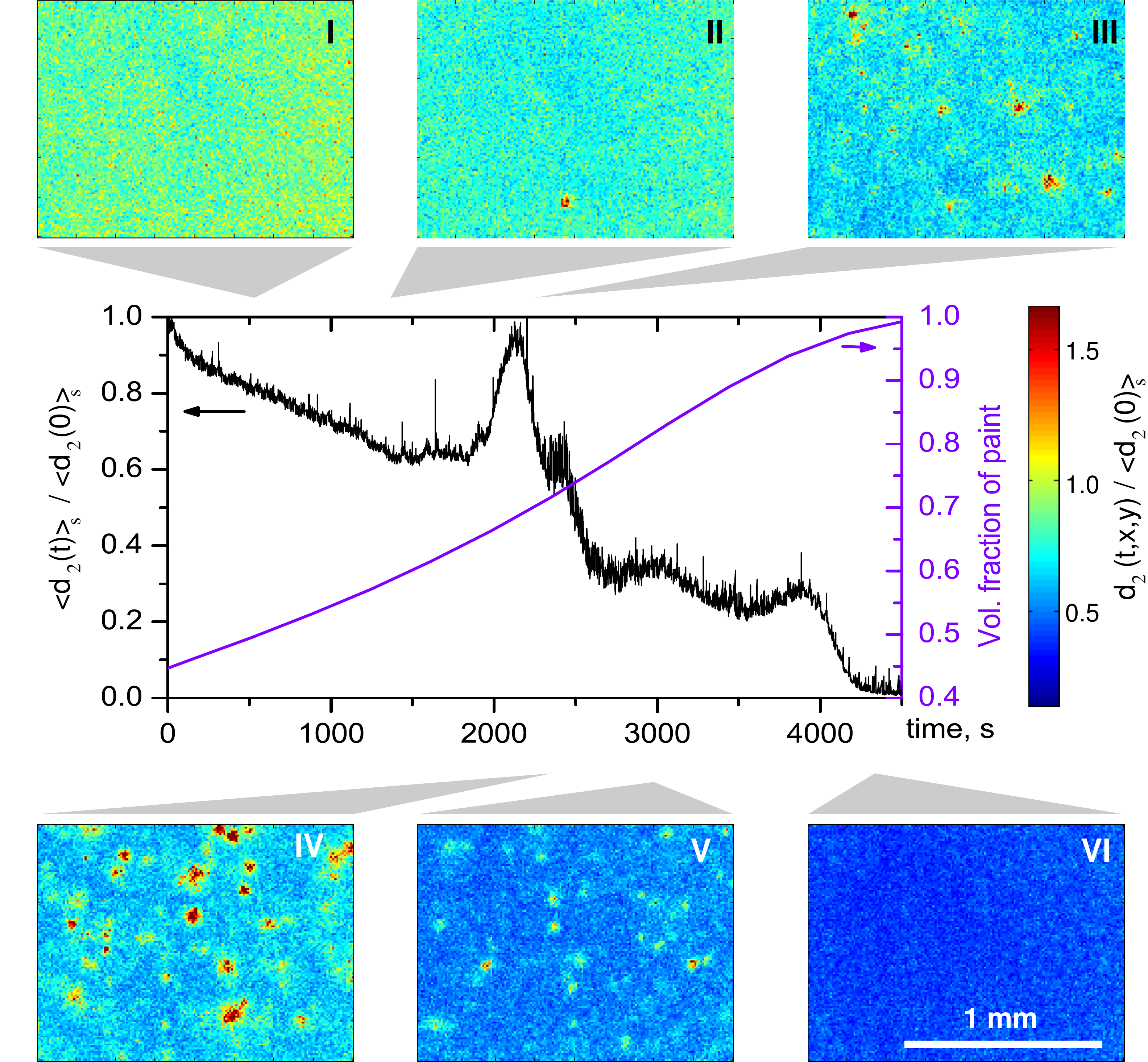}
\vspace{0.1cm} \caption{Dynamic activity analyzed in terms of the normalized intensity structure function (ISF)
$d_2(t, \tau=0.2s)$. The outer panels show ISF maps obtained from \emph{echo speckle imaging} (ESI) at different
stages of the drying process. Center: Time evolution of the ISF $\langle  d_2 (t, \tau= 0.2 sec.) \rangle_s$
measured in the image plane and averaged over the full area of
1.6$\times$1.2 mm$^2$). All data has been normalized with respect to the initial state $t=0$. Solid line: volume fraction of solids in the condensed phase of the film. }
\label{fig:fig2}
\end{figure*}

\section{Imaging local dynamics in a drying paint film}
 The process of colloidal thin film drying can be subdivided in
three stages \cite{scherer:dr}. The first is known as the constant rate period (CRP) of drying. The water
evaporates from the solvent surface as long as the concentration of the liquid phase is high. When the
saturation level falls below the ``critical liquid content'', for monodisperse hard spheres at approximately
64~\% volume fraction, the particles are dynamically arrested. In the subsequent ``first falling rate period''
(FRP1) the drying rate starts to decline. The liquid front now penetrates into the solid bed but water still
remains in a continuous (funicular) phase. While continuously loosing mass the liquid jumps from one
configuration with minimal potential energy to another rapidly withdrawing itself from previously occupied
pores. This process occurs as sudden retreats which are called
``rheon'' events \cite{tsim:scaling99}. Evaporation still occurs near
the surface where the liquid is carried from the inside by fluid flow.  At some point the liquid phase breaks up
into separate fractions (pendular phase) and drying is said to enter the second ``falling rate period'' (FRP2),
completing the drying process.

In our light scattering experiments, during the CRP, we observe the continuous decay of the ISF value with no
spatial or temporal heterogeneities (Fig.~\ref{fig:fig2}, panels I and II). This is the signature of the gradual compaction of
the paint particles and a corresponding increase of viscosity. The drying process of the film enters a new stage
when approaching dynamical arrest. For a volume fraction (condensed phase) of approximately 65~\%, or after
about 2000 seconds, the compaction of the particle structure is stopped. At this transition to the FRP1 the
liquid front retreats into the porous system formed by the paint particles and new solid-air inter-phase
boundaries are formed by uncovered particles. This process provides a new source of refractive index
fluctuations unrelated to the particle motion which manifests themselves as the sudden increase of mean ISF. From
this moment on we directly observe ``rheon'' events in the ISF maps (Fig.2, panels III-V).  As the flux of
liquid to the surface slows down we enter the second falling rate period (FRP2) and the film dries out
completely until $d_2 \left ( t, \tau = 0.2\textrm{ sec} \right ) \rightarrow 0$ (Fig.~\ref{fig:fig2}, panel VI)

\section{Time resolved correlation}
To characterize the intermittency of the signal we have to look in more details on the properties of individual
events and its dependence on the systems size. The standard TRC signal in the image plane can be obtained within the current scheme by calculating the spatially averaged ISF $\langle d_2 (t, \tau) \rangle_s$, keeping the motor at rest (or by replacing the motor with a beam expander). Without the imaging lens we can perform traditional far-field TRC measurements (Figure \ref{fig:fig1}~(b)). For these measurements the ground glass is replaced by a set of lenses and the incident laser beam is then focused onto the sample surface. The camera exposure is set to 4~ms and the time delay is $\tau=0.2$s.


\begin{figure}
\includegraphics[width=1.05\linewidth, trim=10 40 0 0]{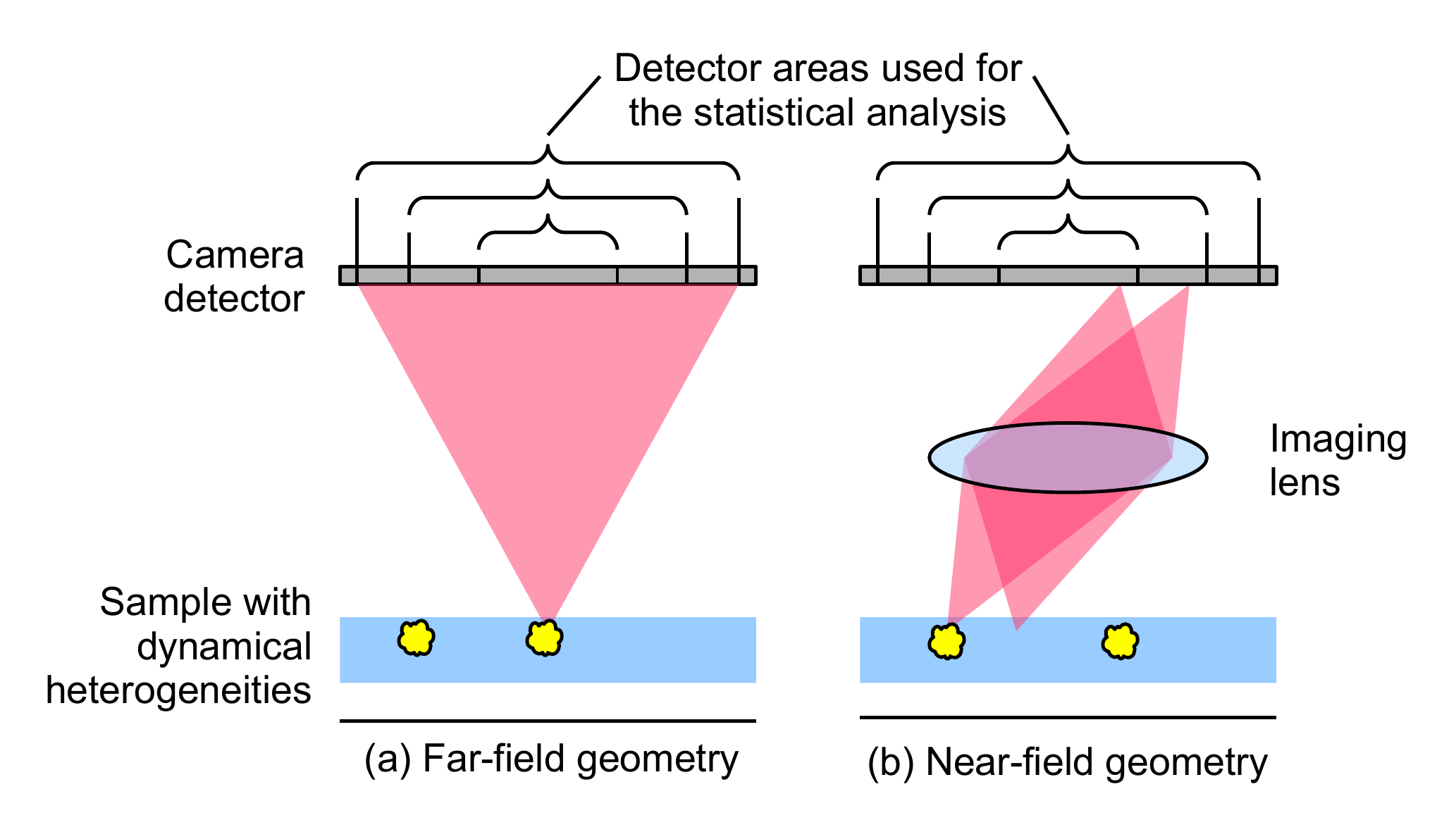}
\vspace{0.1cm} \caption{Time resolved correlation (TRC) experiment in the far-field (left) and in the near field (right).  The selected detector area $A$ sets the number of meta-pixels $N=A/L_0^2$ available for the statistical analysis. For the stationary case  the noise of the intensity structure function scales as  $\sigma_d^2 / \langle d_2 (t, \tau) \rangle^2\simeq 5/N$. The presence of spatially heterogenous dynamics influences both experiments in a distinct manner. a) In the far-field geometry the sample surface area under study remains unchanged when changing $N$. Therefore spatially and temporally heterogenous dynamics leads to a finite value of $\sigma_d^2 / \langle d_2 (t, \tau) \rangle^2$ even for $N \to \infty$ . b) In the near-field \emph{imaging} geometry the sample area increases proportionally to $N$ and therefore a statistical distribution of spatial heterogeneities provides a new source of random fluctuations of $\sigma_d^2 / \langle d_2 (t, \tau) \rangle^2$ albeit with a reduced characteristic number $N'<N$. } \label{fig3a}
\end{figure}

In the following we compare the TRC signal $ \langle d_2(t, \tau) \rangle_s$ both in the image plane and in the far-field (Figure \ref{fig3a}). In the former case we can selectively choose
the sample area monitored by the digital camera. For the latter case we can select the detector area independent of the sample area and can thus compare our data with previous observations of time intermittency in soft glassy materials obtained from similar far-field measurements~\cite{cipelletti05:slow}. A quantitative measure of dynamic heterogeneities is the variance of the ISF $\sigma_d^2 = \langle d_2^2 \left
(t, \tau \right) \rangle - \langle d_2 \left (t, \tau \right ) \rangle^2$ which is the second moment of the
probability density function of $d_2 \left (t, \tau \right )$. In the absence of experimental noise ($N\to \infty$) the variance
tends to zero for a dynamically homogenous system whereas intermittent dynamics leads to a finite residual variance.
Both in the far field and in the near field the quantity $\sigma_d^2$ is however dominated by the presence of
statistical noise of the laser speckle. For the standard Gaussian experimental noise it can be shown that the normalized variance is inversely
proportional to the number $N$ of statistically independent values of  $d_2 (t, \tau)$ recorded by the digital
camera \cite{schatzel:sf}: 

\begin{equation}
\sigma_d^2 / \langle d_2 (t, \tau) \rangle^2 = 5 / N.\label{ISFnoise}
\end{equation}

\begin{figure}
\includegraphics[width=0.7\linewidth]{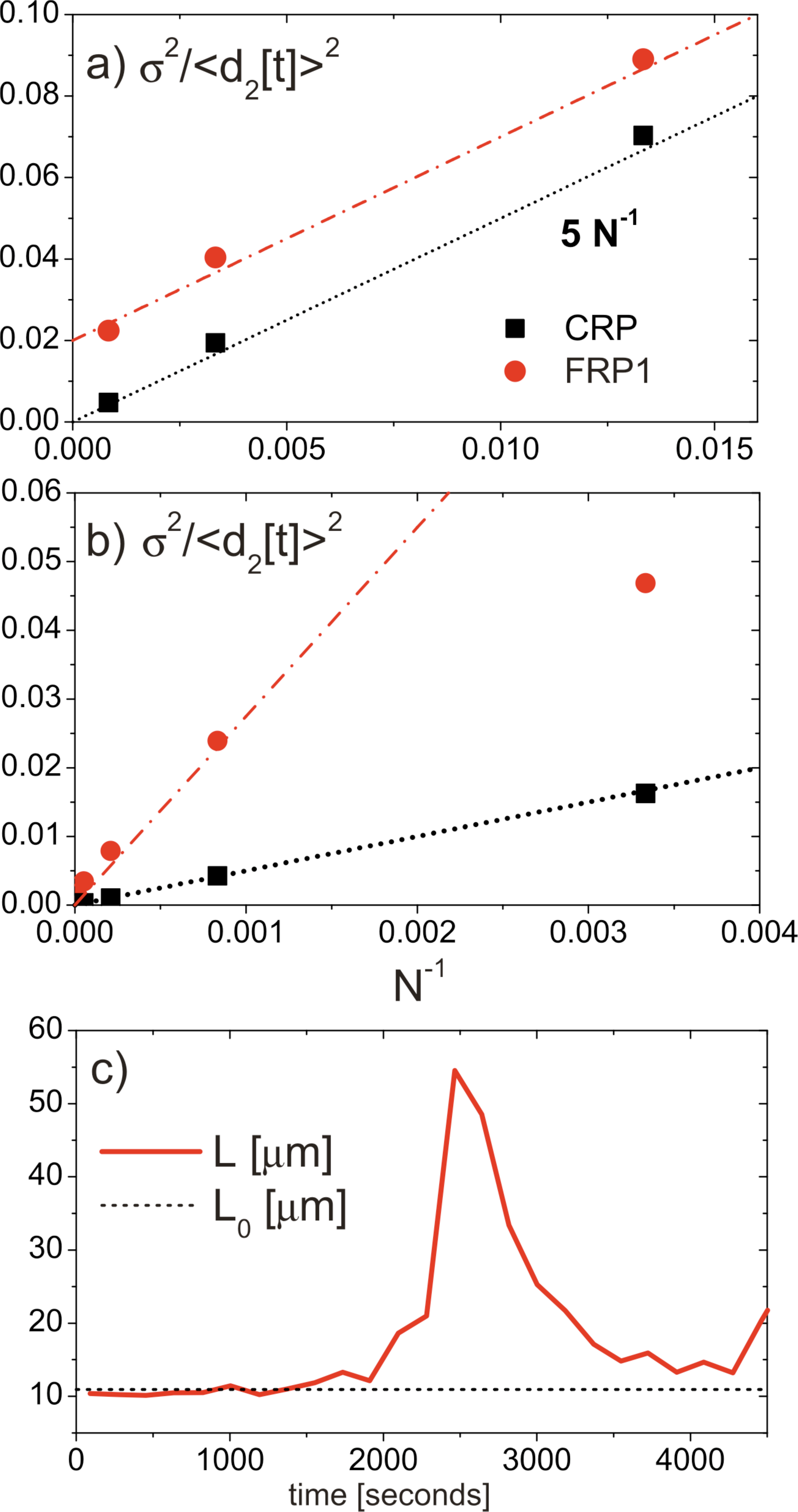}
\vspace{0.1cm} \caption{Statistical properties of the normalized variance of the intensity structure function $\sigma_d^2 / \langle d_2 (t, \tau) \rangle^2$ in the far-field and in the imaging TRC experiment ($\tau=0.2$sec). a) Measurements analyzed in the far field for an incident Gaussian beam of size 220 \textmu m (1/e) during the constant rate period (CRP, $t = 600$~s) and during the first falling rate period (FRP1, $t = 2500$~s) b) Same drying stages observed in the image plane (expanded beam illumination) c) Size of the effective correlated area $L$ as a function of time (for details see text). In the absence of dynamic heterogeneities the lower bound is set by the resolution of the experiment $L_0=11$~\textmu m (dashed line). } \label{fig:fig3}
\end{figure}

For our experimental conditions we expect this relation to be valid in good approximation \cite{Skipetrov2009}. It is thus possible to separate the contribution of intermittent dynamics by analyzing the noise as a function of $N$, which is proportional to the area of the detector matrix $A$ used for the analysis. As suggested by Duri et al. for the far field case the presence of a finite intercept in the limit $1/N \rightarrow 0$ can be exploited to quantify the bare intermittent noise \cite{duri05:trc}. In Figure \ref{fig:fig3}~(a) we demonstrate the intercept in $\sigma_d^2$ originating from intermittent dynamics in the far-field measurement in the FRP1 stage when paint solids occupy about 75~\% of the condensed phase volume \cite{footnote1}.

For measurements in the image plane the situation is different. In the image-plane the speckles are smaller than $L_0$ and the recorded values of $d_2(t, \tau)$ are statistically independent. Thus the number $N$ is directly given by the number of meta-pixels of size $L_0^2$. We note again that in the image plane the number $N$ also scales with the area under consideration. Therefore, if the dynamical heterogeneities are localized in space, increasing $N$ results in an
increasing number of events observed. In this case the sample surface area again contains a
large number of events and Gaussian statistics is recovered according to the central
limit theorem.

Our experiments nicely reveal this feature as shown in Figure \ref{fig:fig3}~(b). During the CRP stage the system dynamics is
homogeneous and the variance $\sigma_d^2$ follows eq.~\eqref{ISFnoise}. In the presence of spatial heterogeneities, and for sufficiently large $N$, the statistical signature is still Gaussian but the slope is now defined by the number of events $N'$ such that $\sigma_d^2 / \langle d_2 \rangle^2 = 5 (N/N') N^{-1}$. We can thus determine $N/N'$
and extract the length scale of the dynamic heterogeneities $L= (N'/N)^{1/2} L_0$. Initially $L$ is equal to the
resolution of our experiment of $L_0 = 11$~\textmu m but in the FRP1 the size grows to approximately 55~\textmu m (Figure \ref{fig:fig3}~(c)). We find this value in good agreement with the direct observations in the imaging experiment
(Fig.\ref{fig:fig2}, Panels II-V). When the size of the area under study becomes comparable to the size of the dynamic
heterogeneity the quantity $\sigma_d^2 / \langle d_2 \rangle^2$ is not well described by Gaussian statistics any
more. In the FRP1 significant deviations from the linear scaling are found for $N^{-1} > 0.001$ or a sample area smaller than $N'*(L_0)^2 \approx 350\times350$~\textmu m$^2$.


\section{Conclusions}
In conclusion we demonstrate that a combination of multi-speckle illumination and time resolved echo speckle imaging enables us to perform spatially resolved dynamic light scattering in real time. The results presented in this work give important input for a better characterization and modelling of the drying process in
general. Moreover the ESI approach can be applied to study many other dense systems. The length scale of 10 \textmu m that can be probed is small enough to resolve spatial heterogeneities in soft glassy materials. At the same time the penetration depth, and the effective scattering volume, is large compared to size of an individual scatterer. Our approach thus allows to access spatially heterogeneous dynamics of soft materials with sufficient resolution and accuracy. We think this method has the potential to bring major advances in the understanding of the slow dynamics of complex systems and the general phenomena of dynamical arrest and the glass transition.

\acknowledgements Financial support by the Swiss National Science Foundation (project number 200020-117762/111824/109137) and the Marie Curie network Grant No. MRTN-CT2003-504712 is gratefully acknowledged. We would like to thank J\"orn Peuser, Sergey Skipetrov and Roberto Cerbino for interesting discussions. Correspondence and request for
materials should be addressed to F.S.


\begin{thebibliography}{35}
\bibitem{footnote1} In the far field the speckles are larger than $L_0$ and $N$ has been determined from a fit to the data in the CRP (solid squares). We note that only for a beams spot size ($1/e$) well below 500 \textmu m we were able to observe the time intermittency shown in Figure \ref{fig:fig3}~(a) in agreement with our observation in the near field (Figure \ref{fig:fig3}~(b)).
\end{thebibliography}

\end{document}